\documentclass[12pt,a4paper]{article}
\usepackage{graphicx}
\usepackage{epsf}
\usepackage{epsfig}
\usepackage{latexsym}
\usepackage{bm}
\usepackage{amssymb}

    \def\bPhi{\mbox{\boldmath$\Phi$}}

\def\bb {\begin {eqnarray}}
\def\ee {\end {eqnarray}}
\newcommand{\supit}[1]{$^{#1}$}
\begin{document}
	
{\Large On rotational-vibrational spectrum of  diatomic beryllium molecule	}
\vspace{5mm}

\centerline{ \large \it  A.A. Gusev\supit{a}, O. Chuluunbaatar\supit{a,b}, S.I. Vinitsky\supit{a,c},
V.L. Derbov\supit{d},}\centerline{ \large \it  A. G\'o\'zd\'z\supit{e},
P.M. Krassovitskiy\supit{a,f}, I.  Filikhin \supit{g}, A.V. Mitin\supit{h,i,j},}\centerline{ \large \it   L.L. Hai\supit{k}, and T.T. Lua\supit{k}}
\vspace{5mm}

\supit{a} Joint Institute for Nuclear Research, Dubna, Russia \\  e-mail: {gooseff@jinr.ru}\\
\supit{b} Institute of Mathematics, National University of Mongolia, Ulaanbaatar, Mongolia\\
\supit{c}RUDN University, Moscow, Russia, 6 Miklukho-Maklaya st, Moscow, 117198\\
\supit{d}N.G. Chernyshevsky Saratov National Research State University, Saratov, Russia \\ e-mail: {derbovvl@gmail.com}\\
\supit{e} Institute of Physics, University of M. Curie-Sk{\l}odowska, Lublin, Poland\\
\supit{f}Institute of Nuclear Physics, Almaty, Kazakhstan\\
\supit{g}Department of Mathematics and Physics,North Carolina Central University, Durham, NC 27707, USA \\
\supit{h} Moscow Institute of Physics and Technology, Dolgoprudny,
            Moscow Region,   Russia \\ e-mail: {mitin.av@mipt.ru} \\
\supit{i} Chemistry Department, Lomonosov Moscow State University, Moscow, Russia\\
\supit{j} Joint Institute for High Temperatures of RAS,   Moscow, Russia\\
\supit{k} Ho Chi Minh city University of Education, Ho Chi Minh city, Vietnam

	\begin{abstract}
The eigenvalue problem for second-order ordinary differential equation (SOODE) in a finite interval with the boundary conditions of the first, second and third kind is formulated. A computational scheme of the finite element method (FEM) is presented that allows the solution of the eigenvalue problem for a SOODE with the known potential function using the programs ODPEVP and KANTBP 4M   that implement FEM in the  Fortran and  Maple, respectively. Numerical analysis of the solution using the KANTBP 4M program is performed for the SOODE exactly solvable eigenvalue problem.  		
The discrete energy eigenvalues and  eigenfunctions  are analyzed for vibrational-rotational states of the diatomic beryllium molecule solving the eigenvalue problem for the SOODE numerically with the table-valued potential function approximated by interpolation Lagrange and  Hermite polynomials and its asymptotic expansion for large values of the independent variable specified as Fortran function.
The efficacy of the programs is demonstrated by the calculations of twelve eigenenergies of vibrational bound states with the required accuracy, in comparison with those known from literature, and the vibrational-rotational spectrum of the diatomic beryllium molecule.\footnote{Submitted to: Proceedings of SPIE }
 \end{abstract}
	
	
	\section{Introduction}

	The study of mathematical models, describing waveguide problems,  spectral and optical properties of diatomic molecular systems, reduces to the solution of a boundary-value problem (BVP) for an elliptic equation of the Schr\"odinger type
\cite{s,Greene2017}.
After the separation of angular variables, this equation reduces to a second order ordinary differential equation (SOODE)  with variable coefficients and the independent variable belonging to the semiaxis $r\in (0,+\infty)$. In this equation the  potential function is numerically tabulated on a non-uniform grid in a finite interval of the independent variable values \cite{m,Bondybey,Patkowski09}.

	To formulate the BVP on the semiaxis, the potential function should be continued beyond the finite interval using the additional information about the interaction of atoms comprising the diatomic molecule at large distances between them. The leading term of the potential function at large distances is given by the van der Waals interaction,
inversely proportional to the sixth power of the independent variable (internuclear distance) with the constant, determined from theory and experimental data        \cite{d,x,Meshkov}.
	
 Therefore, it is necessary to make an appropriate approximation of the tabulated potential function
 and to match the asymptotic expansion of the potential function with its tabulated numerical values (within the accuracy of their calculation) at a suitable sufficiently large value of the independent variable.

	The present paper is devoted to the development of technique for solving the above class of eigenvalue problems with SOODE using the programs   ODPEVP \cite{ODPEVP} and KANTBP 4M \cite{14}  implementing  FEM \cite{2}
 in  Fortran and Maple,   respectively. The technique is applied to the calculation of rotational-vibrational energy spectrum of diatomic berillium molecule.

	\section{
		Setting of the problem}\label{sher1}
	
	The mathematical model describing the spectral and optical characteristics of molecular systems is formulated as a BVP for the SOODE for the unknown function  $\Phi(r)$ of the independent variable $r\in\Omega[r^{\min},r^{\max}]$:
	\begin{eqnarray}
 	\left({D}-E_{}\right)\Phi_{}(r )= \left(-\frac{1}{r^2}\frac{d}{dr}{r^2}\frac{d}{dr}  + {V}(r)-E\right)\Phi(r)=0.\label{1}
	\end{eqnarray}
	Here ${V}(r)$ is a real-valued function from the Sobolev space ${\cal H}_2^{s\geq 1}(\Omega)$,
	providing the existence of nontrivial solutions obeying the boundary conditions (BCs) of the first (I) (Dirichlet), second (II) (Neumann), or third (III) kind at the boundary points of the interval $r\in[r^{\min},r^{\max}]$
with given   $\mathcal{R}(z^t)$:
	\begin{eqnarray}\label{2}
	\mbox{(I)}:\,\,\Phi(r^t)=0, \,\,\ 
 	\mbox{(II)}:\,\, \lim_{r\to r^t} r^2\frac{d\Phi(r)}{dr}
 	=0, \label{3} \,\,\ 
	\mbox{(III)}: \lim_{r\to r^t} r^2\frac{d\Phi(r)}{dr}{=}
	\mathcal{R}(r^t)\Phi(r^t), \, t{=}\min\,\mbox{or}\max. \label{4}
	\end{eqnarray}
	
	The calculation of the approximate solution ${\Phi}(r){\in} {\cal H}_2^{s\geq 1}(\bar \Omega)$  of the BVP (\ref{1})--(\ref{4}) is executed
 by means of the FEM
using the symmetric quadratic functional \cite{2}
	\begin{equation}
 	 \mbox{\boldmath$\Xi$}(\Phi,E,r^{\min},r^{\max})
	=\mbox{\boldmath$\Pi$}(\Phi,E
 )
\label{xiforb}
	  -\Phi(r^{\max})\mathcal{R}(r^{\max})\Phi(r^{\max})
	 +\Phi(r^{\min})\mathcal{R}(r^{\min})\Phi(r^{\min}),
 \nonumber
 \end{equation}

\begin{equation}
	\mbox{\boldmath$\Pi$}(\Phi,E
 ){=}
	\int^{r^{\max}}_{r^{\min}}\Biggl[
	\frac{d \Phi(r)}{d r}
	\frac{d\Phi(r)}{d {r}} \\+
	\Phi(r)({V}(r){-}E)\Phi(r)
	\Biggr]r^2dr\nonumber.
 \end{equation}

	For the bound-state problem the set of $M$ eigenvalues of the energy $E_m$:
	{$E_{1}\leq E_{2}\leq\ldots\leq E_{M}$} and the corresponding set of eigenfunctions
	$\Phi( r)\equiv\{\Phi_{m}( r)\}_{m=1}^{M}$
	is calculated in the space ${\cal H}_2^2$ for the SOODE (\ref{1}). The functions obey the BCs of the first, second or {third} kind at the boundary points of the interval $r\in[r^{\min},r^{\max}]$
	and the orthonormalization condition
	\begin{eqnarray}\label{5}
	\langle \Phi_{m} |\Phi_{m'}\rangle =\int_{r^{\min}}^{r^{\max}}\Phi_{m}(r)\Phi_{m'}(r)r^2dr=\delta_{mm'}.
	\end{eqnarray}
	
	{ Thus, to solve the discrete spectrum problem on an axis or semiaxis, the initial problem is approximated by the BVP in the finite interval
		$r\in [r^{\min}, r^{\max}]$
		with the BCs of the first, second, or third kind with the {given}  $\mathcal{R}(r^t)$,
 {dependent or independent} of the unknown eigenvalue $E$, and the set of approximated eigenvalues and eigenfunctions is calculated. 		
 		
		\subsection {Reduction to an algebraic problem}
		Let us construct a discrete representation of the solution
		$ \Phi_m( r)$ of the problem
		(\ref{1})--(\ref{4}),
		reduced to the variational functional
		(\ref{xiforb})
		on the finite-element mesh
		 \begin{eqnarray}\label{kosetka}
\Omega^{p}_{h_{j}(r)}[r^{\min},r^{\max}]{=}[r_{0}{=}r^{\min},r_{1}, ...,
r_{np-1},r_{np}{=}r^{\max}].\end{eqnarray}
 		The solution
		$ \Phi_m^h( r)\approx  \Phi_m( r)$ is sought in the form of expansion in basis functions $N_{\mu}^g(r)$ in the interval $r\in\Delta=\cup_{j=1}^n\Delta_j=[r^{\min},r^{\max}]$:
		\begin{eqnarray}\label{nnn}
		\Phi_m^h(r)=\sum_{\mu=0}^{L-1}
		\Phi_{m;\mu}^h N_{\mu}^g(r),\quad  \Phi_m^h(r_l)=\Phi_{m;l}^h,
		\end{eqnarray}
		where $L=pn+1$ is the number of the basis functions $N_{\mu}^g(r)$
		and the desired coefficient
 $\Phi^h_{m;\mu}$
  which at $\mu=l$ are values of the function
  $ \Phi^h_{m}(r)$
 at each node  $r=r_l$ of the mesh $\Omega^{p}_{h_{j}(r)}[r^{\min},r^{\max}]$.
 		The basis functions $N_{\mu}^g(r)$ are piecewise continuous polynomials of the order $p$
		in the corresponding subinterval
$r\in\Delta_j=[r^{\min}_j\equiv r_{(j-1)p},r^{\max}_j\equiv r_{jp}]$ constructed using the Lagrange interpolation polynomials (LIP) or Hermite ones \cite{2}.

		The substitution of the expansion (\ref{nnn}) into the variational functional (\ref{xiforb})
 		reduces the BVP (\ref{1})--(\ref{4})
		to the generalized algebraic problem for the set of the eigenvalues  $E_m$ and the eigenvectors
		${\bPhi}^h_m=\{
		\Phi_{
 			m;\mu}^h
 		\}_{\mu=0}^{L-1}:$
		\begin{eqnarray}
		&&  ( {A} - E_m^h\,
		{B})\bPhi_m^h
		=0.\label{eqp}
		\end{eqnarray}
		Here  {$ {A}$ is the symmetric stiffness matrix and ${B}$ is the the positive definite   symmetric  mass matrix}, both having the dimension
 		$L\times L$,  where $L=\kappa^{\max}(np + 1)$.
 \begin{figure}[htbp]
			 \centerline{\includegraphics[width=0.4\textwidth]{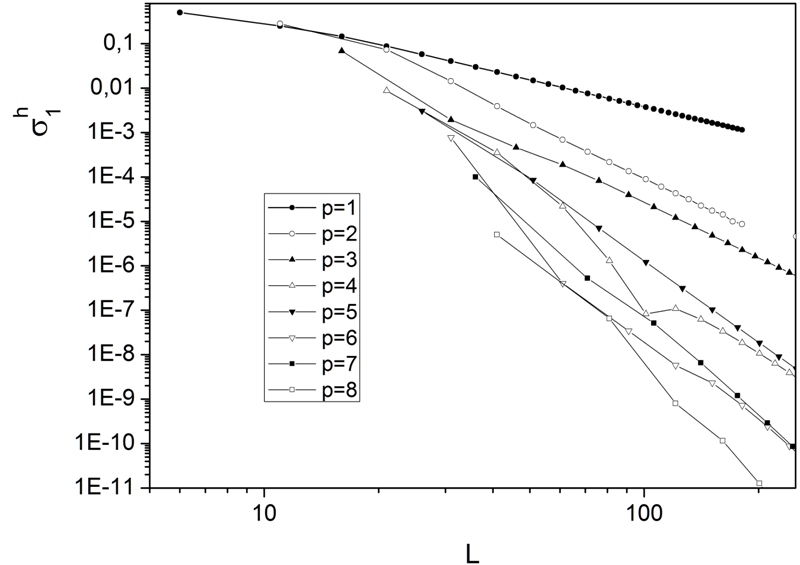}\hfill\includegraphics[width=0.4\textwidth]{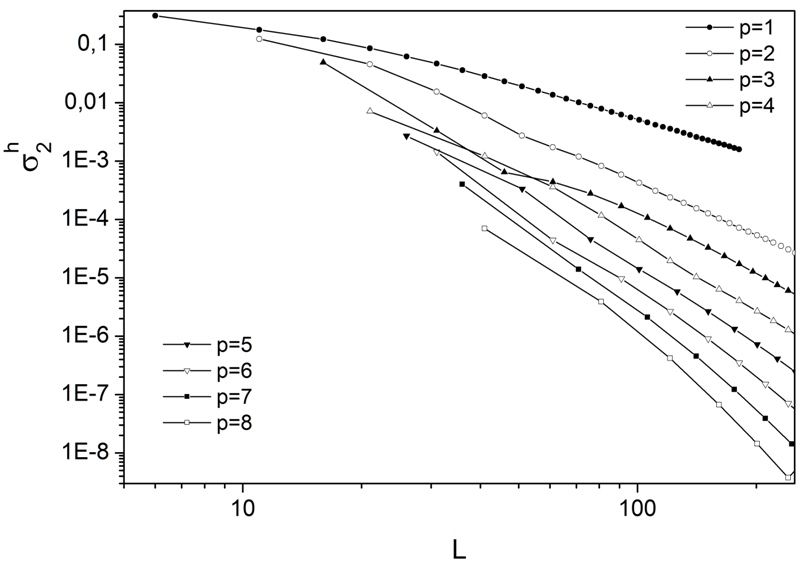}}
\caption{Absolute errors $\sigma_1^h=|E_2^{exact }-E_2^h|$ and
$\sigma^h_2=\max_{z\in\Omega^h(z)}|\Phi_2^{exact}(r)-\Phi_2^{h}(r)|$
				of the second eigenvalue and eigenfunction of the BVP (\ref{1})--(\ref{4})
				as functions of the dimension $L=5n_gp+1$ of the algebraic eigenvalue problem, calculated with the LIP from the first ($p=1$) to the eighth ($p=8$) order.
				The left ends of the curves correspond to the mesh  with one finite element between two nodes. i.e.,  $\Omega =\{0(2n_g)1(2n_g)5(n_g)20\}$ c $n_g=1$,
				where the number $n_g$ of finite elements between two nodes is indicated in parentheses.}
\label{rrr}
 \end{figure}
		
 		Theoretical estimates of the difference between the exact solution $\Phi_m(z)\in {\cal H}^2_{2}$
		and the numerical one
		$\Phi^h_m(r) \in {\bf H^{1}}$
		by the norm ${\bf H^0}$ evaluate the convergence of the eigenvalues and eigenfunctions of the order ${2p}$ and ${p+1}$, respectively \cite{2}:
		\begin{eqnarray}
		\vert  E^h_m   {-}    E_m   \vert \leq
		c_1 h^{2p} ,~~
		\left \Vert \Phi^h_m(r)   {-}\Phi_m(r) \right \Vert_0 {\leq}
		c_2 h^{p+1},\label{theor}
		\end{eqnarray}
		where $h = \max_{1<j<n}h_{j}$ is the maximal step
$h_{j}=r_{j+1}-r_{j}$ of the mesh  (\ref{kosetka}), $c_1\equiv c_1(E_m)>0$ and $c_2\equiv c_2(E_m)>0$ are independent of the step $h$, the norm  ${\bf H^0}$ being defined as
		\begin{eqnarray}\left \Vert \Phi^h_m(r)   {-}\Phi_m(r) \right \Vert_0
		{=}\left (\int\nolimits_{r^{\min}}^{r^{\max}}\!\!\!\!\!r^2dr (\Phi^h_m(r)   {-}\Phi_m(r))^2\right )^{1/2}\!\!\!\!\!\!\!\! .\label{theor1}
		\end{eqnarray}

In the program KANTBP 4M
 the integration
in each finite element is, generally, performed with the potential $V(r)$ approximated by the
   interpolation Hermite polynomials (IHPs) with the node multiplicities  $\kappa^{\max}$, which leads to the quadrature formula \cite{14,2}
 \begin{eqnarray}  \nonumber
 \int_{r_j^{\min}}^{r_j^{\max}} r^2dr N_{L_1}(r,r_j^{\min},r_j^{\max})V(r)N_{L_2}(r,r_j^{\min},r_j^{\max})
\\
 =\sum_{r=0}^p \sum_{\kappa=0}^{\kappa_{\max}-1} V^{(\kappa)}(r_{(j-1)p+r})V_{l_1;l_2;\kappa^{\max}r+\kappa}(r_j^{\min},r_j^{\max})),
 \end{eqnarray}
where $V_{l_1;l_2;l_3}(r^{\min},r^{\max})$ are determined by the integrals with IHPs
 \begin{eqnarray} \nonumber V_{l_1;l_2;l_3}(r_j^{\min},r_j^{\max})
 =\int_{r_j^{\min}}^{r_j^{\max}} r^2N_{l_1}(r,r_j^{\min},r_j^{\max})
  N_{l_2}(r,r_j^{\min},r_j^{\max})N_{l_3}(r,r_j^{\min},r_j^{\max})dr.
 \end{eqnarray}
The obtained expression is exact for polynomial potentials of
the order smaller than $p$. Generally, this decomposition leads
to numerical eigenfunctions and eigenvalues with the accuracy of the order about $p+1 $.

The estimation of the error is carried out using the maximal norm, i.e., the maximal absolute value of the error of the eigenfunctions $\Phi^h_m(r)$ and eigenvalues  $E^h_m$ in the interval $r\in\Omega^h(r)$:
      { \begin{eqnarray}
&&\sigma_{1}=\vert  E^h_m   -    E_m   \vert \leq
		c_1(E_m) \, h^{p+1},
\quad \sigma_{2}=\max_{r\in\Omega^h(r)}|\Phi^h_m({r})-\Phi_m(r)|\leq
		c_2(E_m)  h^{p+1}.\label{theor2}
		\end{eqnarray}}
In the program ODPEVP  the integrals
are calculated using the Gauss
integration rule  with       {$2p + 1$} nodes and  the theoretical estimates (\ref{theor}) hold.

		Since the eigenfunctions of the discrete spectrum exponentially decrease, {$$\Phi^{as}_m(r )\sim\exp (-\sqrt{-E_m}r)/r$$}, at $r\to+\infty$, the initial problem is reduced to a BVP for bound state in the finite interval with the Neumann conditions at the boundary points
		$r^{\min}$ and $r^{\max}$ of the interval and the normalization condition (\ref{5}).
 		
		\begin{figure}[t]
			\centerline{\includegraphics[width=0.48\textwidth]{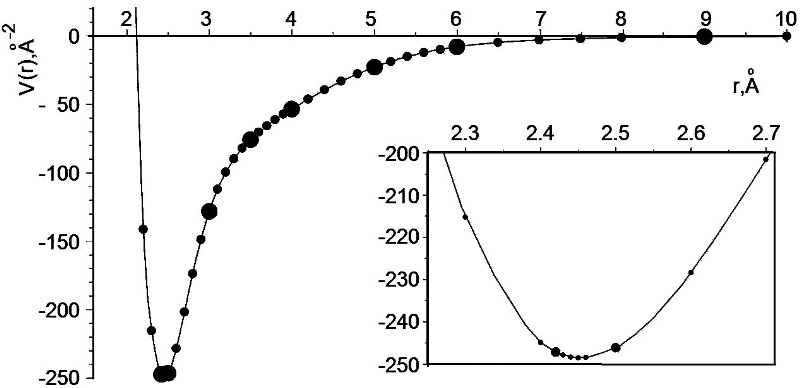}(a)\hfill \includegraphics[width=0.49\textwidth]{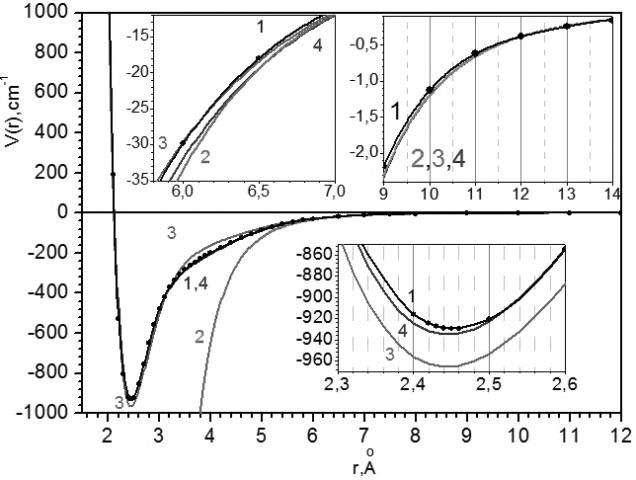}(b)}
			\caption{(a) The potential $V(r)$ (\AA$^{-2}$)  of the beryllium diatomic molecule as a function of $r$ (\AA) obtained by interpolating the tabulated values (points in the subintervals, the boundaries of which are marked by larger-size points) by means of the fifth-order LIPs.
     (b) {The MEMO potential function $V(r)$ (points and line 1  \cite{m}), the
  asymptotic expansion $V_{\rm as}(r)$ of the MEMO function (line 2,  \cite{d}), the analytical forms of the potential function $V_{an}(r)$ (line 3  \cite{x} and  line 4  \cite{Meshkov}). The units for $r$ and $V_*(r)$ are {\AA}  and cm$^{-1}$, respectively.}
}\label{mp}\label{rv6}
\end{figure}
		
		\subsection{
 			Benchmark problem
		}
		
		The original bound state  problem is formulated in the infinite interval ${r}\in(0,+\infty)$ for the Schr\"odinger equation (\ref{1}) with the potential function, inversely proportional to the square of hyperbolic cosine, $V(r)=\frac{-\lambda(\lambda-1)}{\cosh(r)^2}$, where $\lambda>1$.
 The eingenvalues $E^{exact}_m$ and eigenfunctions $\Phi^{exact}_m(r)=r^{-1}\chi^{exact}_m(r)$ of this problem, normalized by the condition (\ref{5}) at $r^{\min}\to 0$ and $r^{\max}\to +\infty$, are known in the analytical form.
 For the chosen $\lambda=11/2$, the
 BVP has two discrete spectrum solutions with the eigenvalues $-E_m=49/4,9/4$.
 		
		The calculations were performed in the finite interval {$r\in[r^{\min},r^{\max}]$} with the Neumann boundary conditions (\ref{3}) on the quasi-uniform mesh
 		{$\Omega =\{0(2n_g)1(2n_g)5(n_g)20\}$},
 		where in parentheses the number of finite elements between two nodes is indicated, the dimension  $L$ is expressed in terms of the number  $n_g$ and the order of LIP $p$ as $L=5n_gp+1$.

		Figure
 \ref{rrr} shows the dependence of the absolute errors (\ref{theor2})
 of
 {the {second} state ($m=2$)} depending on the dimension  $L$ of the algebraic eigenvalue problem (\ref{eqp}) for finite element schemes with LIP of different order  $p$.
		In double logarithmic scale the plots of
 the
 error starting from a certain number $L$ are close to straight lines with different slope, corresponding to the theoretical estimates
 of the approximation order $p+1$ of the approximate eigenfunctions and eigenvalues (\ref{theor2}) using the LIP with different $p$.

\begin{table}[t]
\caption{Eigenvalues of vibrational energy $-E_{vL=0}$ (in  cm$^{-1}$) of  beryllium diatomic molecule calculated using the  programs KANTBP 4M \cite{14} and ODPEVP \cite{ODPEVP} implementing FEM (FEM),
 ab initio MEMO calculation
 \cite{m}, theoretical (EMO) and experimental (Exp) results
 \cite{Bondybey},
 symmetry-adapted perturbation theory (SAPT)\cite{Patkowski09}, and the Morse long-range (MLR) function
 and Chebyshev polynomial expansion (CPE)\cite{Meshkov}.
 $D_e$ is the absolute energy at the dissociation limit in cm$^{-1}$,
 $r_{e}$ is the equilibrium internuclear distance in \AA.
 The 11-th $\chi_{10L}(r)=r\Phi_{10L}(r)$ (solid curves)
 12-th $\chi_{11L}(r)=r\Phi_{11L}(r)$ (dashed curves)
 eigenfunctions vs. $r$ of the vibrational-rotational spectrum of beryllium diatomic molecule at $L=0,1,2$:
 $E_{v=10;L=0}=-4.41$,  $E_{v=10;L=1}=-4.21$,   $E_{v=10;L=2}=-3.82$;
 $E_{v=11;L=0}=-0.325$,   $E_{v=11;L=1}=-0.245$ and $E_{v=11;L=2}=-0.096$ (in cm${}^{-1}$).
 			}\label{t12}
\parbox{0.60\textwidth}{\footnotesize \begin{tabular}{|r|r|r|r|r|r|r|r|r|}\hline
$v$& FEM&MEMO&EMO&Exp&SAPT&{MLR\&CPE}\\\hline
$r_e$&2.4534&2.4534&2.4535&2.4536&2.443&2.445\\\hline
$D_e$&929.804&929.74&929.74&929.7$\pm$2&938.7&934.8\&935.0\\\hline
 0&806.07&806.48&806.5&807.4&812.4&808.1510 \\\hline
 1&583.57&584.32&583.8&584.8&590.1&585.2340 \\\hline
 2&408.73&408.88&408.7&410.3&414.8&410.7319 \\\hline
 3&288.36&288.61&288.3&289.3&292.1&289.7314 \\\hline
 4&211.18&211.42&211.1&212.6&214.5&213.0654 \\\hline
 5&154.16&154.38&154.1&155.9&157.3&156.3536 \\\hline
 6&107.15&107.34&107.1&108.6&109.8&109.1202 \\\hline
 7&68.35 &68.51&68.3&69.7   & 70.7&70.1719 \\\hline
 8&37.80 &37.92&37.7&39.2   & 40.0&39.6508 \\\hline
9&16.33 &16.43&15.8&17.5   & 18.1&17.9772 \\\hline
10& 4.41 &4.40&3.1&4.8      &  5.3&5.3187 \\\hline
11 &0.326 &0.27&    &        &  0.5&0.5175 \\\hline
\end{tabular}
}\hfill
\parbox{0.35\textwidth}{\includegraphics[width=0.35\textwidth,height=0.35\textwidth]{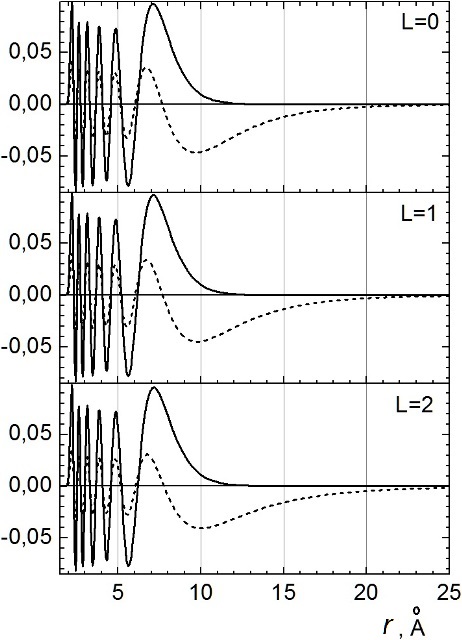}
}
\end{table}

\section{
 			Beryllium diatomic molecule
		}
		In quantum  chemical calculations, the effective potentials of interatomic interaction are presented in the form of numerical tables calculated with limited accuracy and defined on a nonuniform mesh of nodes in a finite domain of interatomic distance values.
However, for a number of diatomic molecules the asymptotic expressions for the effective potentials can be calculated analytically for sufficiently large distances between the atoms.
{The equation for the diatomic
molecules in a crude adiabatic approximation, commonly referred to as Born--Oppenheimer approximation (BO),
has the form}
		\begin{eqnarray}\label{neweq}
		\left({-}\frac{\hbar^2}{2mDa\mbox{\AA}^2}\left(\frac{1}  {r^2}\frac{d}  {dr}{r^2}\frac{d}  {dr}
\right) {+} {\tilde V}_L(\tilde r){-}\tilde E_{vL}\right)\tilde \Phi_{vL}(\tilde r){=}0,
		\end{eqnarray}
where ${\tilde V}_L(\tilde r)={\tilde V}(\tilde r){+}\frac{\hbar^2}{2mDa\mbox{\AA}^2}\frac{L(L{+}1)}{r^2}$,
$L$ is a quantum number of the
total angular momentum,  $\hbar^2/(2Da)$ ${=}$ $1.685762920 \cdot 10^{-7}$ \AA, the reduced
mass of beryllium  is
{$m{=}M/2{=}4.506$},
$\tilde r{=}r$ \AA, the effective potential is
${\tilde V}(\tilde r)$ in atomic units
      {$aue{=}0.002194746314$ \AA$^{-1}$}, the energy is $\tilde E_{vL}$ cm$^{-1}$.

The BVP (\ref{1})--(\ref{4}) was solved for the equation (\ref{neweq})
 		where the variable $r$ is specified in (\AA),
 and the effective potential
 {$V(r)=(2mDa \mbox{\AA}^2 aue/ \hbar^2) {\tilde V}( r\mbox{\AA}){=}58664.99239$ ${\tilde V} ( r\mbox{\AA})$ \AA$^{-2}$},  and the desired value of energy $E_{vL}{=}(2mDa \mbox{\AA}^2 /\hbar^2) \tilde E_{vL}$ in \AA$^{-2}$, $\tilde E_{vL}{=}(1/0.2672973729)E_{vL}$ cm$^{-1}$.

 {In  Ref. \cite{m} the potential $V(r)$ (see {Fig.} \ref{mp}) is
given by the BO-PRC potential function marked as MEMO tabular values $\{V^{M}(r_i)\}_{i=1}^{76}$.
So, in the interval $r\in [r_1=1.5,r_{46}=9]$ the potential $V(r)$ was approximated  in subintervals $r\in [r_{5k-4}, r_{5k+1}]$, $k=1,...,9$ by the fifth-order interpolation Lagrange polynomials of the variable $r$.
In the interval $r\in[r_{\rm match}=14,\infty)$ the asymptotic behavior $V_{\rm as}(r){=}58664.99239 \tilde V_{\rm as}(r)$ at large $r$ is expressed as   \cite{d}
\begin{equation}\label{dd}
\tilde V_{\rm as}(r){=}{-}\left(\frac{214(3)}{Z^6}{+}\frac{10230(60)}{Z^8}{+}\frac{504300}{Z^{10}}\right),\\
\end{equation}
where $Z=r/0.52917$.
In the subinterval $r\in[r_{46}=9,r_{\rm match}=14]$ we consider the approximation of  the potential $V(r)$
by the fourth-order interpolation Hermite polynomial
using the  values of the potential $V(r)$ at the points $r=\{r_{46}=9,r_{47}=10,r_{48}=11\}$ and the values of the asymptotic potential $V_{\rm as}(r)$ and its derivative $d V_{\rm as}(r)/dr$ at the point $r=r_{\rm match}=14$.
This approximation is  specified in \AA$^{-2}$ as  REAL*8 FUNCTION VPOT(R) of the variable $R$ in (\AA) (see Appendix).
 		
{ For comparison,  {Fig.} \ref{rv6}  plots the above potential function $V(r)$, its asymptotic expansion $V_{\rm as}(r)$, and the analytical potential functions $\tilde V_{an}(r)$ in a.u. proposed in Ref. \cite{x}:
\begin{eqnarray}\nonumber
\tilde V_{an}(r){=}A\exp({-}bZ) {+}d\exp({-}eZ{-}fZ^2){-}\sum_{n{=}3}^8\left(\left(1{-}\exp({-}bZ)\sum_{k{=}0}^{2n} \frac{(bZ)^k}{k!}\right)\frac{C(2n)}{Z^{2n}}\right),\label{fdd}
		\end{eqnarray}
		where
		$A{=}21.7721$,
		$b{=}1.2415$,
		$d{=}-4.3224$,
		$e{=}0.5891$,
		$f{=}0.0774$,
		$Z{=}r/0.52917$,
		$C(6) {=}214$, $C(8) {=} 10 230$, $C(10) {=} 504 300$,
		$C(2i){=}(C(2i-2)/C(2i-4))^3C(2i-6)$, $i{=}6,7,8$, and $r$ is given in \AA.
One can see that  the MEMO potential function $V(r)$  has a minimum $-D_e({\rm FEM}){=}V(r_{e})=929.804$cm$^{-1}$  at the equilibrium point $r_e=2.4534$       {\AA} and  displaces above the analytic potential function $V_{\rm an}(r)$ in the vicinity of this point, $-D_e({\rm Sheng}){=}V_{\rm an}(r_{e}){=}{-}948.3$cm$^{-1}$  and the MLR\&CPE potential functions \cite{Meshkov} $-D_e({\rm MLR}){=}934.8$, $-D_e({\rm CPE}){=}935.0$ at $r_{e}=2.445$,
while the analytical potential function $V_{\rm an}(r)$ is located above the MEMO and MLR\&CPE
potential functions in the interval $r\in(3.2,6.1)$,
i.e. to the left of the interval $r\in(6.1,\infty)$, where the considered potentials tend
to the dominated asymptotic potential $V_{as}(r)$.

In the calculation presented below, we used
the asymptotic expansion $V_{\rm as}(r)$, Eq. (\ref{dd})
		with which the matching of the tabulated potential $V (r)$
 and the  asymptotic   potential  $V_{\rm as}(r)$
 was executed at  $r=r_{\rm match}=14$ using REAL*8 FUNCTION VPOT(R) of the variable $R$ in (\AA) (see Appendix).
The BVP (\ref{1})
 		was solved on the finite element mesh
		$\Omega_{1}=\{$1.50 ($n_g$) 2.00 ($n_g$) 2.42 ($n_g$) 2.50 ($n_g$) 3.00 ($n_g$) 3.50 ($n_g$) 4.00 ($n_g$) 5.00 ($n_g$) 6.00 ($n_g$) 9.00 ($n_g$) 14.00 ($n_g$) 19.00 ($n_g$) 24.00 ($n_g$) 29.00 ($n_g$) 38.00 ($n_g$) 48.00 ($6n_g$) ${r^{\max}}=$78.00$\}$ with Neumann BCs. In each of the subintervals (except the last one) the potential $V(r)$ was approximated by the LIP of the fifth order, and $n_g=4$ finite elements were used.
 The last integrand was divided into $6n_g$ finite elements and the potential  $V(r)$ was replaced with its asymptotic expansion. In the solution of the BVP at all finite elements of the mesh the local functions were represented by the fifth-order LIP.

Table  \ref{t12} presents the results of using FEM programs KANTBP 4M and ODPEVP to calculate {twelve energy eigenvalues
of beryllium} diatomic molecule.
Note, that our calculation was performed using the program that implements the Numerov method
on the mesh (0,100) for twelve levels
with the mesh spacing $0.02$ with Dirichlet BCs for $\chi_{vL}(r)=r\Phi_{vL}(r)$,
which differs from the FEM results in Table \ref{t12} only in the last significant digit.
The table shows
the eigenvalues calculated with ab initio modified (MEMO) expanded Morse oscillator (EMO) potential function  \cite{m}. In contrast to the original EMO function, which was used to describe the experimental (Exp) vibrational levels  \cite{Bondybey}, it has not only the correct dissociation energy, but also describes all twelve vibrational energy levels with the RMS error smaller than 0.4 cm$^{-1}$. }

{ The table also shows the results of recent calculation using the Morse long-range (MLR) function and Chebyshev polynomial expansion (CPE) alongside with the EMO potential function \cite{Meshkov}. Similar results have been obtained in Ref. \cite{prz2018}.  The main attention in the optimization of the MLR and CPE functions was
focused on their correct long-range behavior displayed in Fig. \ref{rv6}.
 However, there are some problems with the quality of the MLR and CPE potential curves \cite{m}.
 As a consequence, one can see from the table, that the MLR and CPE  results provide
 {\it  a  lower estimate} while FEM and MEMO results
 give {\it an upper estimate}
 for the discrete spectrum of the diatomic beryllium  molecule.}
		\begin{figure}[htbp]
			\centerline{\includegraphics[width=0.48\textwidth]{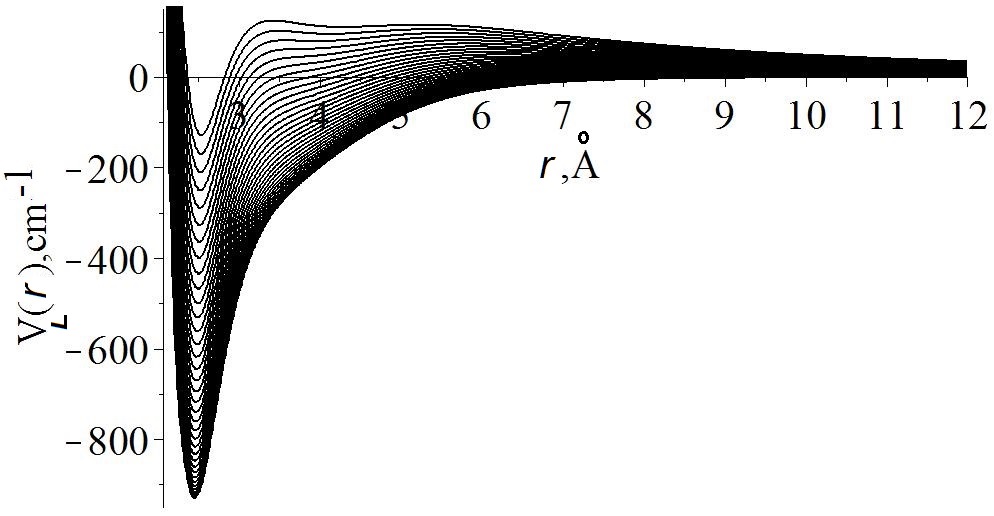}\hfill
\includegraphics[width=0.48\textwidth]{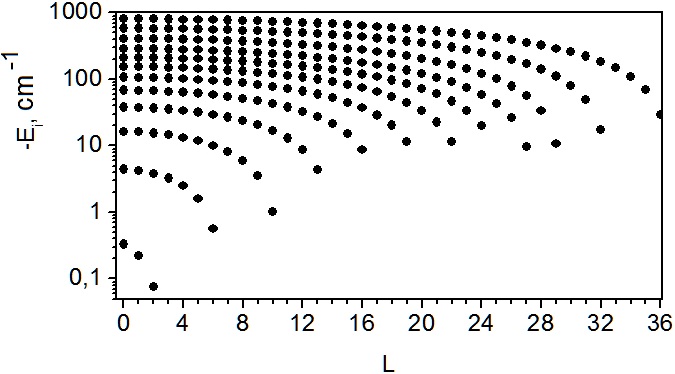}}
			\caption{Potential functions $V_L(r)$ (in cm$^{-1}$) vs $r$ (in \AA) at  $L=0,...,36$ and rotational-vibrational spectrum $E_{vL}$ (in cm$^{-1}$) of the beryllium diatomic molecule vs $L$.
 			}\label{bel12}
		\end{figure}

 Figure \ref{bel12}  displays the potential functions
 $V_L(r)$ from $L=0$ till $L=36$ that support
$36+33+30+28+25+23+20+17+14+11+7+3+1=248$ vibrational--rotational levels or
$12+12+12+11+11+11+11+10+10+10+10+9+9+9+8+8+8+7+7+7+6+6+6+5+5+4+4+4+3+3+2+2+2+1+1+1+1=248$
 rotational-vibrational levels.
Figure \ref{bel12}  also shows  the rotational-vibrational spectrum $E_{vL}$ (in cm$^{-1}$)
of the beryllium diatomic molecule vs $L$.
One can see that potentials $V_L(r)$ at $L=1$ and $L=2$ supports 12 vibrational energy levels.
Figure 1 (right) shows the behavior of the 11-th $\chi_{10L}(r)=r\Phi_{10L}(r)$
and  12-th $\chi_{11L}(r)=r\Phi_{11L}(r)$ eigenfunctions  of the vibrational-rotational spectrum of beryllium diatomic molecule at $L=0,1,2$.

	\section*{Conclusion}
 	We present the computational finite element scheme for the solution of the BVP for the SOODE with variable coefficients using the programs  KANTBP 4M and ODPEVP.
	The numerical analysis of the solution of the benchmark eigenvalue problem for the SOODE is given.

The discrete energy eigenvalues and eigenfunctions  are analyzed for
vibrational--rotational states of the diatomic beryllium molecule
by  solving the eigenvalue problem for the SOODE numerically with the table-valued potential function
 approximated by interpolation Lagrangian and  Hermite polynomials and its asymptotic expansion for large values of the independent variable specified as Fortran function.

	      The efficacy of the programs is demonstrated by the calculations of twelve eigenenergies of the vibrational bound states of the diatomic beryllium molecule with the required accuracy in comparison with those known from literature, as well as the vibrational-rotational spectrum.

New high accuracy \textit{ab initio} calculations of the tabulated potential function will be useful
for further study of 
the vibrational-rotational spectrum and scattering problems.

The results and the presented FEM programs with interpolation Hermite polynomials that preserve the derivatives continuity   of the approximate solutions  can be  {applied}
in the analysis of  spectra of diatomic molecules and waveguide problems  by solving the eigenvalue and scattering problems in the closed--coupled channel method.

The work was partially supported by the RFBR (grants No. 16-01-00080 and No.18-51-18005), the Bogoliubov-Infeld program, the Hulubei-Meshcheryakov program, the RUDN University Program 5-100,  grant of Plenipotentiary of the Republic of Kazakhstan in JINR, and Ho Chi Minh city University of Education (grant CS.2018.19.50). 
 \appendix
 {\scriptsize\sc
 \begin{verbatim}
      REAL*8 FUNCTION VPOT(R)                               #-279.496863252389085774320D0
      REAL*8 R                                              #+16.3002196867918408302045D0*(R-3.50D0)**5
      IF ( R .LT. 0.200D1) THEN                             #+60.2807956755861261238267D0*(R-3.50D0)**3
      VPOT = -25773.7109044290317516659D0*R                 #-47.2260081876105825000000D0*(R-3.50D0)**4
     #+45224.0477977149109075999D0                          #-47.2940696984941697748181D0*(R-3.50D0)**2
     #+11630.1409366263902691980D0*(R-1.50D0)**5             ELSEIF ( R .LT. 0.500D1) THEN
     #-21410.9944579041874319967D0*(R-1.50D0)**3             VPOT =37.5433740382779941025814D0*R
     #-6655.69301415296537793622D0*(R-1.50D0)**4            #-203.532767180257088384326D0
     #+37646.6374905803929755811D0*(R-1.50D0)**2            #+1.57933446805903445309567D0*(R-4.00D0)**5
      ELSEIF ( R .LT. 0.242D1) THEN                         #+2.06536720797980643219389D0*(R-4.00D0)**3
      VPOT = -3104.29731660146789925758D0*R                 #-3.90913978185322013018878D0*(R-4.00D0)**4
     #+6567.96677835187237746414D0                          #-6.72175482977376484481239D0*(R-4.00D0)**2
     #+145901.637557436977844389D0*(R-2.00D0)**5             ELSEIF ( R .LT. 0.600D1) THEN
     #+70890.0501932798636244675D0*(R-2.00D0)**3             VPOT = 22.4749425088812799926950D0*R
     #-178091.891722217850289831D0*(R-2.00D0)**4            #-135.176802468861661924475D0
     #-5215.01465840480348371833D0*(R-2.00D0)**2            #-1.74632723645421934973176D0*(R-5.00D0)**5
      ELSEIF ( R .LT. 0.250D1) THEN                         #-1.13910625636659500903584D0*(R-5.00D0)**3
      VPOT = -87.4623224249792247412537D0*R                 #+3.46551546383436915312500D0*(R-5.00D0)**4
     #-35.4722493028120322541662D0                          #-8.23297658169947934799179D0*(R-5.00D0)**2
     #-5122452.98252855176907985D0*(R-2.42D0)**5             ELSEIF ( R .LT. 0.900D1) THEN
     #-37267.2557427451395256506D0*(R-2.42D0)**3             VPOT = 8.25446369250102969043326D0*R
     #+767538.576723810368564874D0*(R-2.42D0)**4            #-57.5068241812660846645295D0
     #+1940.26376259904725429059D0*(R-2.42D0)**2            #+0.262554831228989391666862D-1*(R-6.00D0)**5
      ELSEIF ( R .LT. 0.300D1) THEN                         #+1.52595797003340802069435D0*(R-6.00D0)**3
      VPOT =95.5486415932416181588138D0*R                   #-.302331762133111382686686D0*(R-6.00D0)**4
     #-485.009680038558034254034D0                          #-4.49546478165576415777827D0*(R-6.00D0)**2
     #-559.791178882855959174489D0*(R-2.50D0)**5             ELSEIF ( R .LT. 0.1400D2) THEN
     #-2399.49666698491294179656D0*(R-2.50D0)**3             VPOT = 11.385941234992376680396136937226D0*R
     #+2045.36781464380745875587D0*(R-2.50D0)**4            #-37.683304037819782698889968642231D0
     #+1039.16158144865749926292D0*(R-2.50D0)**2            #-1.3036988112705401175758401661849D0*R**2
      ELSEIF ( R .LT. 0.350D1) THEN                         #+0.6675467548036330733418010128614D-1*R**3
      VPOT = 181.680445623994493034163D0*R                  #-0.12861577375486918213137485397657D-2*R**4
     #-673.209766066617684340488D0                           ELSE
     #-42.1375729651804208384176D0*(R-3.00D0)**5             Z=R/0.52917D0
     #+154.527717281912104793036D0*(R-3.00D0)**3             VPOT = -( 214.D0/Z**6+10230.D0/Z**8
     #-2.64421453511851874413350D0*(R-3.00D0)**4            # +504300.D0/Z**10)
     #-224.971347436273044312400D0*(R-3.00D0)**2             ENDIF
      ELSEIF ( R .LT. 0.400D1) THEN                          VPOT =58664.99239D0*VPOT
      VPOT = 58.2170634592331667146628D0*R                   RETURN
\end{verbatim}
}

\end{document}